\documentclass[graybox]{svmult}

\usepackage{type1cm}        
\usepackage{makeidx}         
\usepackage{graphicx}        
\usepackage{multicol}        
\usepackage[bottom]{footmisc}

\usepackage{adjustbox}

\usepackage{listings,lstautogobble}

\usepackage{color,soul}
\usepackage{url}
\usepackage{booktabs}

\usepackage{savesym}
\usepackage{amsmath}
\savesymbol{iint}
\usepackage{txfonts}
\restoresymbol{TXF}{iint}

\usepackage{caption}

\makeindex             

\usepackage{natbib}
\begin{document}

\title*{Semantically-Enriched Search Engine for Geoportals: A Case Study with ArcGIS Online}
\titlerunning{Semantically-Enriched Search Engine for ArcGIS Online}
\author{Gengchen Mai \and Krzysztof Janowicz \and Sathya Prasad  \and  Meilin Shi \and Ling Cai \and Rui Zhu \and Blake Regalia \and  Ni Lao}
\authorrunning{G. Mai et al.}
\institute{Gengchen Mai \at STKO Lab, UC Santa Barbara \\ \email{gengchen_mai@geog.ucsb.edu}
\and Krzysztof Janowicz \at STKO Lab, UC Santa Barbara \\ \email{jano@geog.ucsb.edu}
\and Sathya Prasad \at ESRI Inc, Redlands, CA, USA \\ \email{sprasad@esri.com}
\and Meilin Shi \at STKO Lab, UC Santa Barbara \\ \email{meilinshi@ucsb.edu}
\and Ling Cai \at STKO Lab, UC Santa Barbara \\ \email{lingcai@ucsb.edu}
\and Rui Zhu \at STKO Lab, UC Santa Barbara \\ \email{ruizhu@geog.ucsb.edu }
\and Blake Regalia \at STKO Lab, UC Santa Barbara \\ \email{blake.regalia@gmail.com}
\and Ni Lao \at SayMosaic Inc. \\ \email{noon99@gmail.com}
}
%
%
\maketitle

\vspace*{-2.2cm}
\abstract{
Many geoportals such as ArcGIS Online are established with the goal of improving geospatial data reusability and achieving intelligent knowledge discovery. However, according to previous research, most of the existing geoportals adopt Lucene-based techniques to achieve their core search functionality, which has a limited ability to capture the user's search intentions. To better understand a user's search intention, query expansion can be used to enrich the user's query by adding semantically similar terms. In the context of geoportals and geographic information retrieval, we advocate the idea of semantically enriching a user's query from both geospatial and thematic perspectives. In the geospatial aspect, we propose to enrich a query by using both place partonomy and distance decay.
In terms of the thematic aspect, concept expansion and embedding-based document similarity are used to infer the implicit information hidden in a user's query. This semantic query expansion framework is implemented as a semantically-enriched search engine using ArcGIS Online as a case study. A benchmark dataset is constructed to evaluate the proposed framework. Our evaluation results show that the proposed semantic query expansion framework is very effective in capturing a user's search intention and significantly outperforms a well-established baseline -- Lucene's practical scoring function -- with more than 3.0 increments in DCG@K (K=3,5,10).
}


\begin{center}
	\begin{figure}
		\centering
		\includegraphics[width=1.0\textwidth]{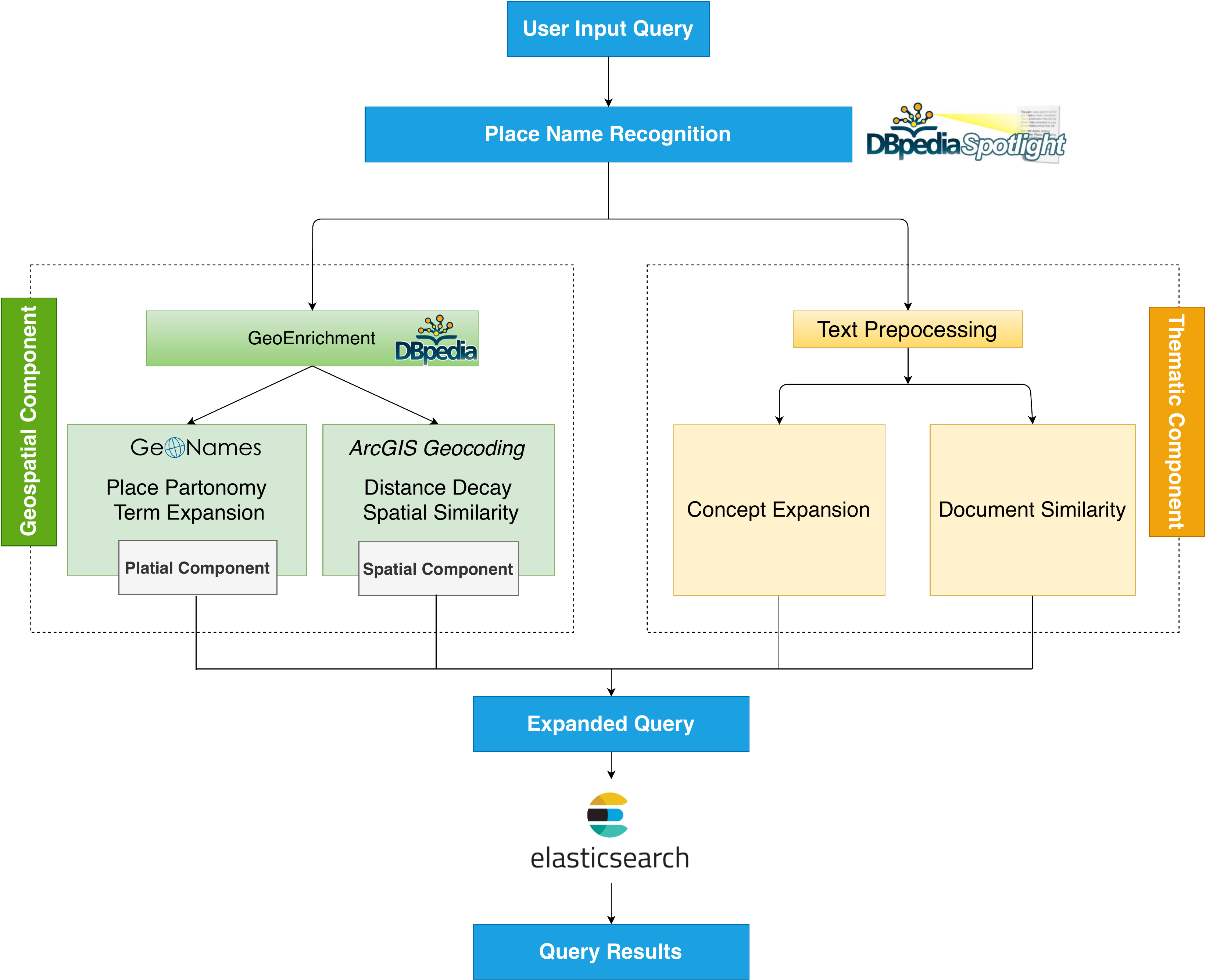}
		\caption{The semantic query expansion framework
		}
		\label{fig:workflow}
	\end{figure}
\end{center}

\section{Introduction} \label{sec:intro}

The increasing growth of geospatial data poses a great challenge to data discovery, access, and maintenance \citep{jiang2018towards}. In order to increase data reusability and facilitate geospatial knowledge discovery, many geoportals have been established to provide integrated access to  geospatial resources \citep{hu2015enabling}. Examples of geoportals include the DataOne 
Data Catalog\footnote{\url{https://search.dataone.org/data}}, U.S. Geological Survey Science Data Catalog\footnote{\url{https://data.usgs.gov › datacatalog}}, NASA Earth data Search\footnote{\url{https://search.earthdata.nasa.gov/search}}, ArcGIS Online, and so on.

The most important component of a geoportal is its search functionality, which is usually supported by geographic information retrieval (GIR) techniques. Generally speaking, information retrieval (IR) aims at finding relevant entries based on a user's query. The entries can be documents, websites, services, maps, and so on, depending on the application scenarios. As a subfield of IR, geographic information retrieval \citep{jones2008geographical} adds space (and time) as additional dimensions to the traditional information retrieval problems \citep{janowicz2011semantics}. In addition to traditional thematic similarity, spatial (and temporal) similarity is considered when the relevance score between a user's query $q$ and an entry $d$ is calculated. 

Despite the success of GIR in academia, in practice, the core search functionality of most existing geoportals is still based on 
Apache Lucene or Elasticsearch \citep{jiang2018towards}. These Lucene-based engines 
use a term frequency-inverse document frequency (TF-IDF) approach to compute the similarities between a user's query and document entries, which is insufficient to completely capture a user's search intention. For example, when a user searches for \textit{natural disaster in California} (Query $q_{1}$), (s)he is probably more interested in a document which describes the Kincade Fire that burned in Sonoma County on Oct. 23rd, 2019 since wild fires are a type of natural disaster and Sonoma County is a subdivision of California. However, if this document contains neither the term ``natural disaster'' nor ``California'', a Lucene-based model will give a zero relevance score between this document and the Query $q_{1}$, thus resulting in a low recall. This highlights the necessity of understanding the user's search intentions both semantically and spatially in a (G)IR system.

 According to \citet{dominich2008modern}, IR can be formally defined as: 
 \begin{equation}
IR = m[\mathcal{R}(D, (q, \langle I, \mapsto \rangle))]
\label{equ:ir}
\end{equation}
 where $m$ is the degree of relevance; $\mathcal{R}$ is the relevance relationship; $D$ is a set of (document) entries; $q$ is the user's query; $I$ and $\mapsto$ are implicit and inferred information. The most challenging part in this equation is the question of how to obtain the implicit and inferred information $I$, $\mapsto$ based on user queries. \textit{Query expansion} techniques, which add terms and conditions to a user query with the goal of improving the query-object relevance score \citep{vechtomova2009query}, can be utilized to semantically take the user's search intention into account.

The traditional query expansion  focuses on semantically-enriching a user's query from a thematic perspective. In the context of geoportals (e.g., ArcGIS Online) we argue that a user's query should be expanded (or semantically-enriched) from two perspectives: thematic and geospatial. 
In the thematic aspect, a query can be enriched/expanded by adding thematically similar concepts/terms. For example, as for Query $q_{1}$, some highly related topics of ``natural disaster'' such as \textit{earthquake}, \textit{wild fire}, \textit{flood}, and \textit{hurricane} can be added to the original query. 
In a geoportal, extra attention should be paid to the geospatial aspect. Geospatially related terms can be added to the query. For example, as for Query $q_{1}$, we can consider adding the names of the subdivisions of California to the query. Since this process relies on the place hierarchy, we call it \textit{platial} query expansion. Moreover, the spatial scopes of the query and entries can also be used to compute the spatial similarity between them. 
After being enriched/expanded from these two perspectives, the new query is applied to the geoportal in the hope of improving the recall of the GIR system.

Note that the core idea of query expansion is to minimize the mismatch between a user query and candidate entries so that the recall of the IR system is improved. A similar idea can be applied when we calculate spatial similarities between a user's query and entries. Most of the traditional spatial similarity measures are based on topological relations between the spatial scopes of the user's query and an entry. For example, \citet{jiang2018towards} defined the spatial similarity between a query $q$ and a document entry  $d$, denoted as $Sim(q, d)$, based on their geographic scopes $Area(q)$, $Area(d)$ as well as their intersection $Area(q \cap d)$ (See Equation \ref{equ:spa_sim}).
\begin{equation}
Sim(q, d) = \Big (  \dfrac{Area(q \cap d)}{Area(q)} + \dfrac{Area(q \cap d)}{Area(d)}\Big  ) \times 0.5
\label{equ:spa_sim}
\end{equation}

According to Equation \ref{equ:spa_sim}, if $Area(q \cap d) = 0$, then $Sim(q, d) = 0$ which means if the intersection of the geographic footprints of $q$ and $d$ is zero, the spatial similarity score is zero. 
This may lead to a loss of valuable spatial proximity information
in many scenarios. 
To give a concrete example, if a user searches for \textit{Weather in Los Angeles} (Query $q_{2}$), a map $d_{1}$ about \textit{Temperature in Oxnard} should be considered more relevant than, say,  $d_{2}$ which is about \textit{Temperature in Southern Africa}. However, since the both geographic scopes of Oxnard and Southern Africa do not intersect with the footprint of Los Angeles ($Area(q_{2} \cap d_{1}) = 0$ and $Area(q_{2} \cap d_{2}) = 0$), we will have $Sim(q_{2}, d_{1})=0$ and $Sim(q_{2}, d_{2})=0$ according to Equation \ref{equ:spa_sim} which does not match our intuition.

In other words, it might be better to utilize a distance decay function here instead and minimize the mismatch between the current query $q_{2}$ and $d_{1}$. Inspired by this observation, we utilize a Gaussain kernel distance decay function to compute the spatial similarity between the spatial scopes/geographic footprints between the query and documents. Using a distance decay function to optimize the query-document relevance is also related to work on query relaxation in the context of geographic question answering \citep{mai2019relaxing}.

\textbf{The research contributions of this work are as follows:}
\begin{enumerate}
    \item We propose a semantic query expansion framework for geoportals which enriches a user's query from both thematic and geospatial aspects.
    \item We develop a semantically-enriched search engine prototype for ArcGIS Online by implementing the proposed query expansion framework. 
    \item We collect a benchmark dataset to evaluate the presented framework against a widely used baseline model - Lucene's practical scoring function. The evaluation results show that our semantic query expansion framework outperforms the baseline by a significant margin.
\end{enumerate}

The remainder of this work is structured as follows. 
In Sec. \ref{sec:relatedwork}, several work about geographic information retrieval are discussed. Next, we present our query expansion framework and describe each component of this system in Sec. \ref{sec:method}. Particularly in Sec. \ref{subsec:data} we discuss about the reproducibility of our work and provide guidelines related to data sets and software that facilitate future research along this line.
In Sec. \ref{sec:exp}, we introduce a benchmark dataset we collect to evaluate our GIR framework and then discuss the evaluation results.
Finally in Sec. \ref{sec:conclusion} we conclude our work and discuss the future research directions.
\vspace*{-0.5cm}
\section{Related Work} \label{sec:relatedwork}

The idea of query expansion is to reformulate a user's query by adding semantically related concepts \citep{azad2019query} to minimize the query-object mismatch and increase the recall of an IR system. This typically comes at the expense of reducing the precision. Generally speaking, query expansion techniques can be classified into two categories: global analysis and local analysis \citep{azad2019query}. As for global analysis, the expansion terms are selected based on manually built knowledge bases, knowledge graphs, or large corpora. Finding semantically related terms based on word embedding \citep{mikolov2013distributed,mai2018combining} or topic modeling \citep{hu2015metadata} is an example. Local analysis refers to query expansion methods that select expansion terms based on the retrieved documents of the initial user's query. Example models include relevance feedback \citep{rocchio1971relevance} and pseudo-relevance feedback \citep{buckley1995automatic}. 
In this work, we adopt the global analysis method and use word embedding to select semantically related terms of query terms. 

Many query expansion techniques are not directly applicable for geospatial terms. For example, it is more reasonable to select geospatially related terms based on place hierarchies (e.g., from a digital gazetteer) rather than using word embedding models. This suggests a need for separately handling  geospatial aspect in a query expansion task. For instance, \citet{huang2008hierarchical} classified queries into two types - location sensitive and location non-sensitive - and then handled them by using different query expansion techniques. 

In the field of geographic information retrieval, there are a few works aiming at ranking documents based on both textual and spatial relevance such as the multi-dimensional scattered ranking method proposed by \citet{van2005multi}. Our work follows a similar research direction but also add \textit{platial} similarity to the ranking algorithm.

In addition to query expansion, another line of work for building a semantically-enriched search engine for geoportals is to enrich the metadata. For example, \citet{hu2015enabling} converted the metadata of ArcGIS Online items into Linked Data and then enriched the metadata to enable semantic search. Similar to our idea, \citet{hu2015enabling} also considered the semantic enrichment in two aspects: thematic and geospatial. However, converting data into another format for semantic enrichment requires additional processing steps, storage, and maintanance to keep both data sources in sync. In this work, we focus on enabling semantic search by using query expansion techniques in which the underlying data storage (e.g., Elasticsearch, Apache Lucene) remains unchanged.
\vspace*{-0.5cm} 
\section{Method} \label{sec:method}
In this section, we will first describe the dataset and project setup in Section \ref{subsec:data}. Next, we describe our semantic query expansion framework in detail. The proposed framework is composed of two major components - geospatial component and thematic component - which focus on different aspects. Figure \ref{fig:workflow} shows the overall architecture of the proposed framework. We will present each component below with the example query \textit{Chicago traffic} (Query $q_{3}$).


\vspace*{-0.5cm}
\subsection{Data and Software Availability}  
\label{subsec:data}
Developed by Environment System Research Institute (ESRI), ArcGIS Online is one of the best-known web geoportals. It contains a collections of web maps, data layers, tools, services, and applications contributed from different GIS users all over the world \citep{hu2015enabling}. 
Elasticsearch\footnote{\url{https://www.elastic.co/}}, a widely used search and analytic engine, is utilized to store the metadata of these ArcGIS Online items and support the portals searching functionality. 
The metadata of each ArcGIS Online item has different fields such as ``id'', ``title'', ``snippet'', ``description'', ``type'', ``location'' (point), ``coordinates'' (the bounding box) and so on. 
The core search functionality of ArcGIS Online is based on Lucene's 
query-document similarity function which is computed based on term frequency and inverse document frequency (TF-IDF) scoring such as Lucene's practical scoring function
\footnote{\scriptsize{\url{https://www.elastic.co/guide/en/elasticsearch/guide/2.x/practical-scoring-function.html}}}, Okapi BM25, and so on. 
Therefore, Lucene's practical scoring function is a natural baseline for our semantic query expansion framework.

In order to establish an evaluation dataset for our search engine prototype, we collect 53,404 items using the ArcGIS Online RESTful API which contains 
1) all items published by Esri or its related organizations before September 2017; 
2) all items published on ArcGIS Online between June and September in 2014 and 2017. 

We use Elasticsearch to host all the retrieved ArcGIS Online items. The proposed semantic query expansion framework will serve as a middle layer as shown in Figure \ref{fig:workflow} to semantically-enrich the current user query. The expanded query will be sent to the established Elasticsearch index to get relevant ArcGIS Online items. The motivation here is to enable semantic search functionality on top of a portal such as ArcGIS Online without changing the underlying layers, e.g., data storage. In order to evaluate the proposed semantic query expansion framework and compare it with the baseline, namely Lucene's practical scoring function, we also conduct a human participant test to get query-document relevance scores through Amazon Mechanical Turk sandbox\footnote{\url{https://www.mturk.com/}}. Detail description about this benchmark dataset can be found in Section \ref{subsec:eval}. The data and source code are available at \footnote{\url{https://github.com/gengchenmai/arcgis-online-search-engine}} including 1) the evaluation benchmark dataset; 2) the source code of our query expansion framework. 
The established database is hosted by Elasticsearch 5.4.0\footnote{\url{https://www.elastic.co/blog/elasticsearch-5-4-0-released}} with a vector scoring plugin\footnote{\url{https://github.com/MLnick/elasticsearch-vector-scoring}} to enable word embedding computation.

\subsection{Query Preprocessing: Place Name Recognition}

\hspace*{-\parindent*2}%
\begin{minipage}[c]{\textwidth}
	\begin{lstlisting}[basicstyle=\ttfamily, captionpos=b, caption={An example of DBpedia SPARQL query generated in the GeoEnrichment step to get more information about the identified places.}, label={lst:dbq},frame=single]
	select ?place ?lat ?long ?label ?area ?geoid {
	OPTIONAL {
	    ?place geo:lat ?lat.
	    ?place geo:long ?long.
	}
	OPTIONAL {
	    ?place rdfs:label ?label.
	    FILTER(lang(?label) = "en")
	}
	OPTIONAL {
	    ?place <http://dbpedia.org/ontology/PopulatedPlace/areaTotal> ?area.
	}
	OPTIONAL {
	    ?place owl:sameAs ?geoid .
	    FILTER(CONTAINS(str(?geoid), 'geonames'))
	}
	VALUES ?place {
	    <http://dbpedia.org/resource/Chicago>
	}
	} 
	\end{lstlisting}
\end{minipage}

Given a query such as \textit{Chicago traffic}, we need to first split it into a geospatial aspect and a thematic aspect. A place name recognition service (e.g., DBpedia Spotlight\footnote{\url{https://www.dbpedia-spotlight.org/}}) is utilized to recognize the toponyms appearing in the query (in this case the city of Chicago) and then link it to the corresponding entities (\texttt{dbo:Chicago}) in a knowledge graph such as Wikidata or DBpedia. The identified places are then handled by the geospatial query expansion component and the rest of the query is send to the thematic query expansion component.

\vspace*{-0.5cm} 
\subsection{Geospatial Query Expansion Component} \label{subsec:geo}
The geospatial query expansion component focuses on improving the \textit{platial} and \textit{spatial} similarity between a user's query and a candidate ArcGIS Online item. 

In order to facilitate the following query expansion process, we first enrich the identified geographic entities with additional information such as geographic coordinates, place names, total area, and their GeoNames identifier (See Listing \ref{lst:dbq}). We call this GeoEnrichment step (See Figure \ref{fig:workflow}).

\subsubsection{Platial Component}  \label{subsec:platial}
The platial component focuses on finding similar geographic terms based on the place hierarchy. We use the GeoNames\footnote{\url{https://www.geonames.org/}} service to get the top $K$ subdivisions of the identified places. For example, we can add \textit{Belmont Cragin} and \textit{Englewood} as expanded geographic terms to the expanded query of Query $q_{3}$. Here, the platial similarity between a query $q$ and an ArcGIS item $d_{o}$, denoted as $Sim_{platial}(q,d_{o})$, is defined as 
\begin{equation}
Sim_{platial}(q,d_{o}) = \sum_{p_{i} \: in \: q} W_{geo}(p_{i}) \sum_{p_{ij} \: \in \: Q_{platial}(p_{i}) \cup \{p_{i}\}} W_{platial}(p_{i},p_{ij}) \sum_{f_{k} \: in \: d_{o}}  W_{f}(f_{k}) M(p_{ij},f_{k})
\label{equ:platial}
\end{equation}
Here $p_{i}$ refers to the $i$th identified place from $q$; $W_{geo}(p_{i})$ is the relative importance of place $p_{i}$ among all the identified places and $\sum_{p_{i} \: in \: q} W_{geo}(p_{i}) = 1$; $Q_{platial}(p_{i})$ refers to the set of expanded geographic terms; $W_{platial}(p_{i},p_{ij})$ indicates the importance of $p_{ij} \in Q_{platial}(p_{i}) \cup \{p_{i}\}$ with respect to the corresponding place $p_{i}$; $W_{f}(f_{k})$ indicates the weight of matching one specific metadata field $f_{k}$ since matching some fields such as ``title'' is much more important than matching other fields such as ``description'' and $\sum_{f_{k} \: in \: d_{o}} W_{f}(f_{k}) = 1$; $ M(p_{ij},f_{k})$ indicates the number of matches of the expanded geographic term $p_{ij}$ in the current field $f_{k}$. 

\subsubsection{Spatial Component}  \label{subsec:Spatial}
The spatial component measures the spatial similarity between a query $q$ and item $d_{o}$. \citet{frontiera2008comparison} discussed different geometric approaches to accessing spatial similarity and most of them are computed based on the topological relationships between the geographic scopes of query $q$ and item $d_{o}$. An example of similarity measures is Jaccard similarity index \citep{jaccard1912distribution}. Some non-topological relation based spatial similarity indices also exist such as Hausdorff Distance. 
 
In this work, we use a distance decay approach with Gaussian kernels. Each identified place has a Gaussian kernel which is placed at the center of its bounding box. The bandwidth of a kernel is determined based on the bounding box of the corresponding place. The intuition comes from Tobler's First Law of Geography: the relatedness between query $q$ and item $d_{o}$ decreases with respect to their distance. Here ArcGIS Geocoding API is utilized to obtain the bounding boxes of the identified places. The spatial similarity $Sim_{spatial}(q,d_{o})$ is defined in Equation \ref{equ:spatial} where $Gauss(p_{i}, d_{o})$ is the Gaussian score between identified place $p_{i}$ and item $d_{o}$. The impact of different spatial similarity measures on the performance of this semantic query expansion framework will be left for future work.

\begin{equation}
Sim_{spatial}(q,d_{o}) = \sum_{p_{i} \: in \: q} W_{geo}(p_{i}) Gauss(p_{i}, d_{o})
\label{equ:spatial}
\end{equation}

\vspace*{-0.5cm} 
\subsection{Thematic Query Expansion Component} \label{subsec:thematic}
As the name indicates, thematic query expansion focuses on minimizing the query-item mismatch from a thematic, i.e., topic-based, point of view. To achieve this, we adopt two approaches: concept expansion and embedding-based document similarity. We will discuss each of them below.

Before performing thematic query expansion, some text preprocessing steps such as tokenization, word lemmatization, and stop word removal have been taken to extract thematic concepts/terms from the user's query such as \textit{natural}, \textit{disaster} in Query $q_{1}$ and \textit{traffic} in Query $q_{3}$.

\subsubsection{Concept Expansion Component} \label{subsubsec:w2v}
The idea of concept expansion is to find thematically similar terms to the query terms and add them to the expanded query clause. This is a common way to do query expansion \citep{jiang2018towards,hu2015metadata}. Unlike the previous work in GIR which use semantic knowledge base \citep{jiang2018towards} or topic modeling \citep{hu2015metadata} to find thematically similar terms, we use word embedding technique \citep{mikolov2013distributed} to achieve this. A similar approach has been used in developing academic search engine \citep{mai2018combining}. Given the term \textit{traffic}, word embedding model finds thematically similar terms such as \textit{congestion, rail, train, roads}, and so on.

Equation \ref{equ:w2v} shows the thematic similarity between $q$ and $d_{o}$ based on concept expansion $Sim_{concept}(q,d_{o})$. Here, $t_{i}$ indicates a thematic term in the user's query such as \textit{traffic}. $W_{thematic}(t_{i})$ means the normalized weight of $t_{i}$ among all thematic query terms and $\sum_{t_{i} \: in \: q} W_{thematic}(t_{i}) = 1$. $T_{w2v}(t_{i})$ indicates the set of thematically similar terms of $t_{i}$ based on a pretrained word embedding model such as GLove \citep{pennington2014glove} and $W_{w2v}(t_{i}, t_{ij}) = \dfrac{cosine(t_{i}, t_{ij})}{\sum_{t_{ix} \: \in \: T_{w2v}(t_{i}) \cup \{t_{i}\}} cosine(t_{i}, t_{ix})}$ indicates normalized weight of term $t_{ij}$ with respect to $t_{i}$ based on their cosine similarity. $M(t_{ij},f_{k})$ refers to the number of matches of the expanded thematic term $t_{ij}$ in the current field $f_{k}$.

\begin{equation}
Sim_{concept}(q,d_{o}) = \sum_{t_{i} \: in \: q} W_{thematic}(t_{i}) \sum_{t_{ij} \: \in \: T_{w2v}(t_{i}) \cup \{t_{i}\}} W_{w2v}(t_{i}, t_{ij}) \sum_{f_{k} \: in \: d_{o}} W_{f}(f_{k}) M(t_{ij},f_{k})
\label{equ:w2v}
\end{equation}

\subsubsection{Embedding-Based Document Similarity Component} \label{subsubsec:d2v}
Instead of explicitly matching the expanded thematic terms to ArcGIS Online items, the embedding-based document similarity compares query $q$ and item $d_{o}$ in the hidden word embedding space. Equation \ref{equ:d2v} shows how the similarity score is defined. $E_{query}(q) = \sum_{t_{i} \: in \: q} Word2Vec(t_{i})$ is the embedding of query $q$ which is computed by simply adding the word embeddings of each thematic terms in the query $q$. $E_{doc}(d_{o})$ is the document embedding of $d_{o}$ which is computed based on TF-IDF weighted word embedding of each terms in its title, snippet, and description. 

\begin{equation}
Sim_{doc}(q,d_{o}) = cosine \Big ( E_{query}(q), E_{doc}(d_{o}) \Big)
\label{equ:d2v}
\end{equation}

\subsection{Expanded Query Construction} \label{subsec:expandquery}
The overall similarity between a query $q$ and an ArcGIS Online iterm $d_{o}$ is a weighted sum of all four components: platial (place-based) component, spatial component, concept expansion component, and embedding-based document similarity component. $\lambda_{platial}$, $\lambda_{spatial}$, $\lambda_{concept}$, and $\lambda_{doc}$ are their corresponding weights.

\begin{multline}
Sim(q,d_{o}) =  \lambda_{platial} * Sim_{platial}(q,d_{o}) + \lambda_{spatial} * Sim_{spatial}(q,d_{o}) + \\ \lambda_{concept} * Sim_{concept}(q,d_{o}) + \lambda_{doc} * Sim_{doc}(q,d_{o})
\label{equ:sim}
\end{multline}

\begin{center}
	\begin{figure}
		\centering
		\includegraphics[width=1.0\textwidth]{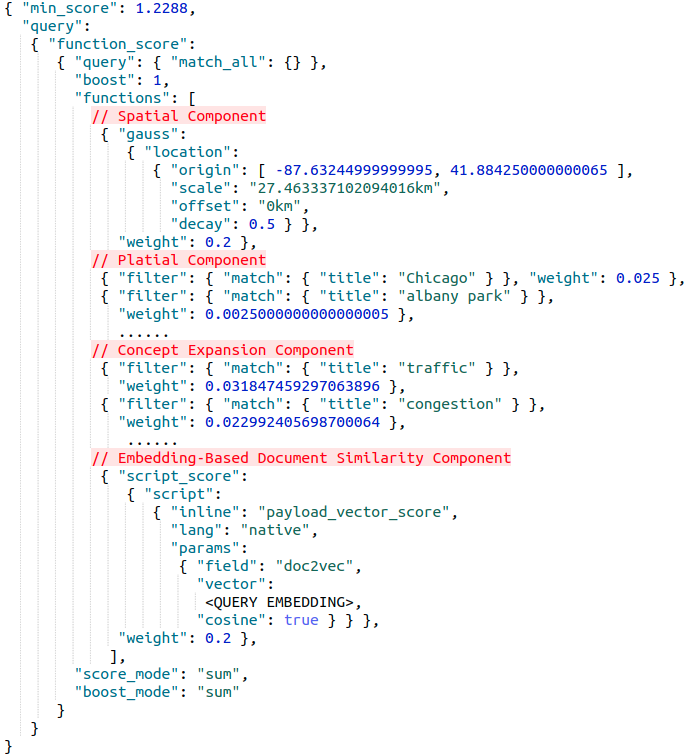}
		\caption{An example of Elasticsearch query derived from our semantic query expansion framework.} \label{fig:es_query}
		\vspace*{-0.7cm}
	\end{figure}
\end{center}

In practice, each component can be written as a collection of function score query clauses in Elasticsearch. Figure \ref{fig:es_query} shows an example of Elasticsearch query constructed after the proposed semantic query expansion framework for the given \textit{Chicago traffic} query. Each component is highlighted. Executing this expanded query in the established Elasticsearch index will give us the final search result.

\vspace*{-0.7cm}

\vspace*{-0.5cm}
\section{Experiment} \label{sec:exp}

\subsection{Semantically-Enriched Search Engine}   \label{subsec:interface}
\begin{center}
	\begin{figure}
		\centering
		\includegraphics[width=1.0\textwidth]{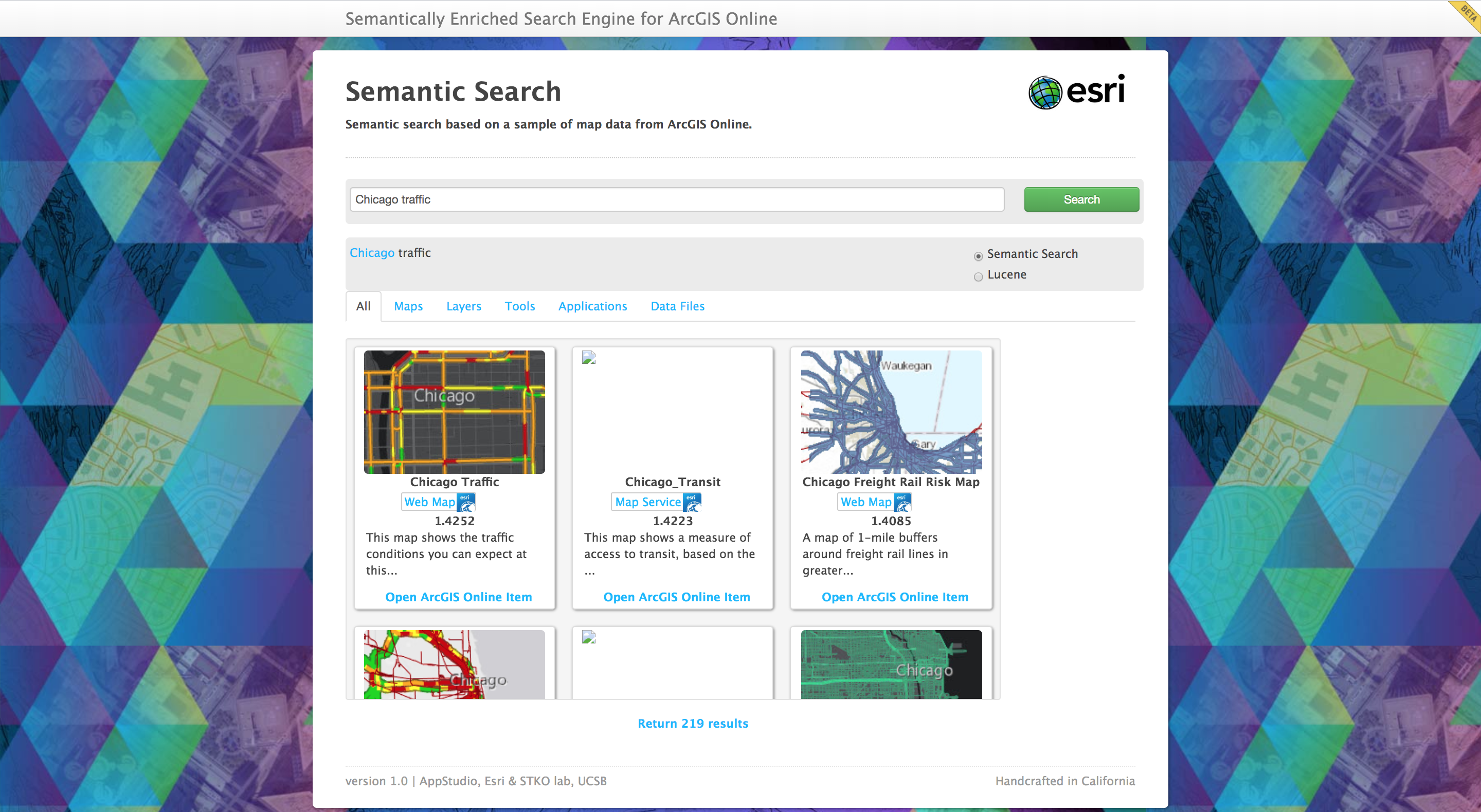}
		\caption{A web interface for the semantically-enriched search engine prototype for ArcGIS Online.} \label{fig:interface}
		
	\end{figure}
\end{center}

\begin{figure}
	\centering
	\begin{minipage}{.5\textwidth}
		\centering
		\includegraphics[width=0.9\linewidth]{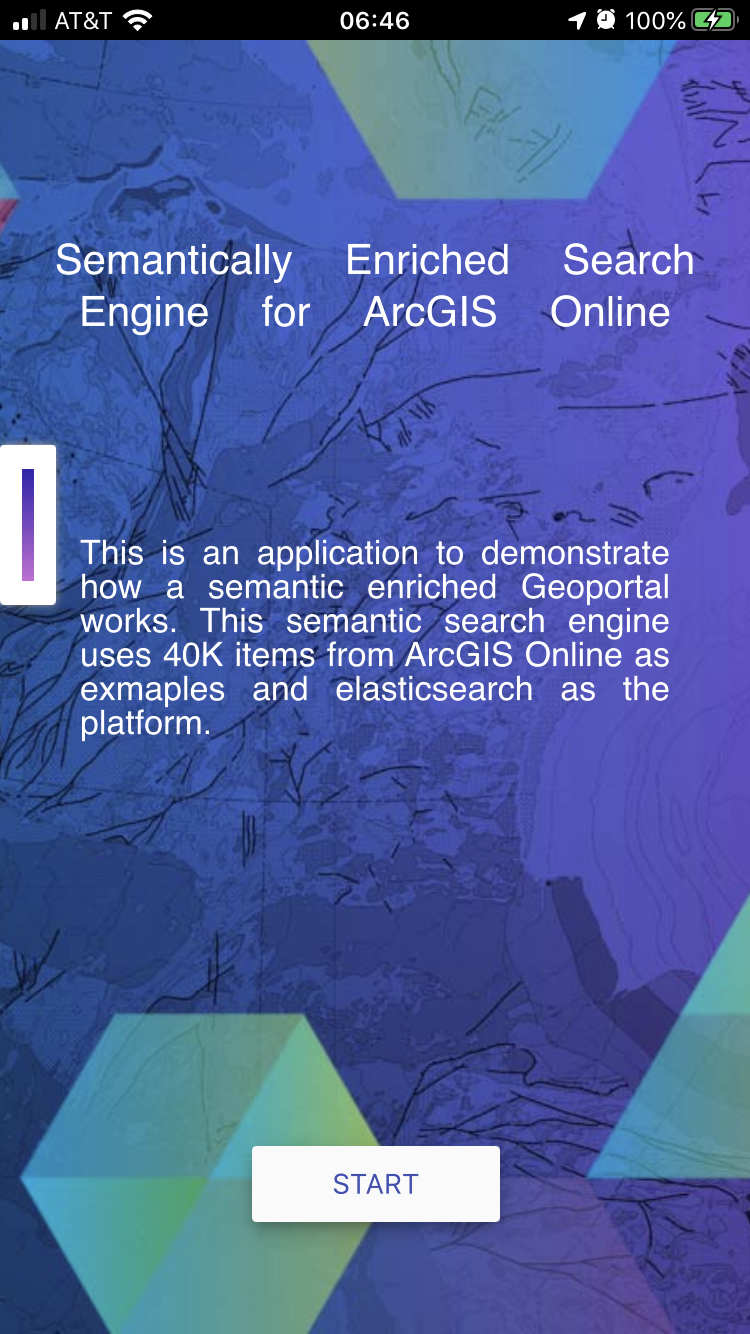}
		\label{fig:mobile_start}
	\end{minipage}%
	\begin{minipage}{.5\textwidth}
		\centering
		\includegraphics[width=0.9\linewidth]{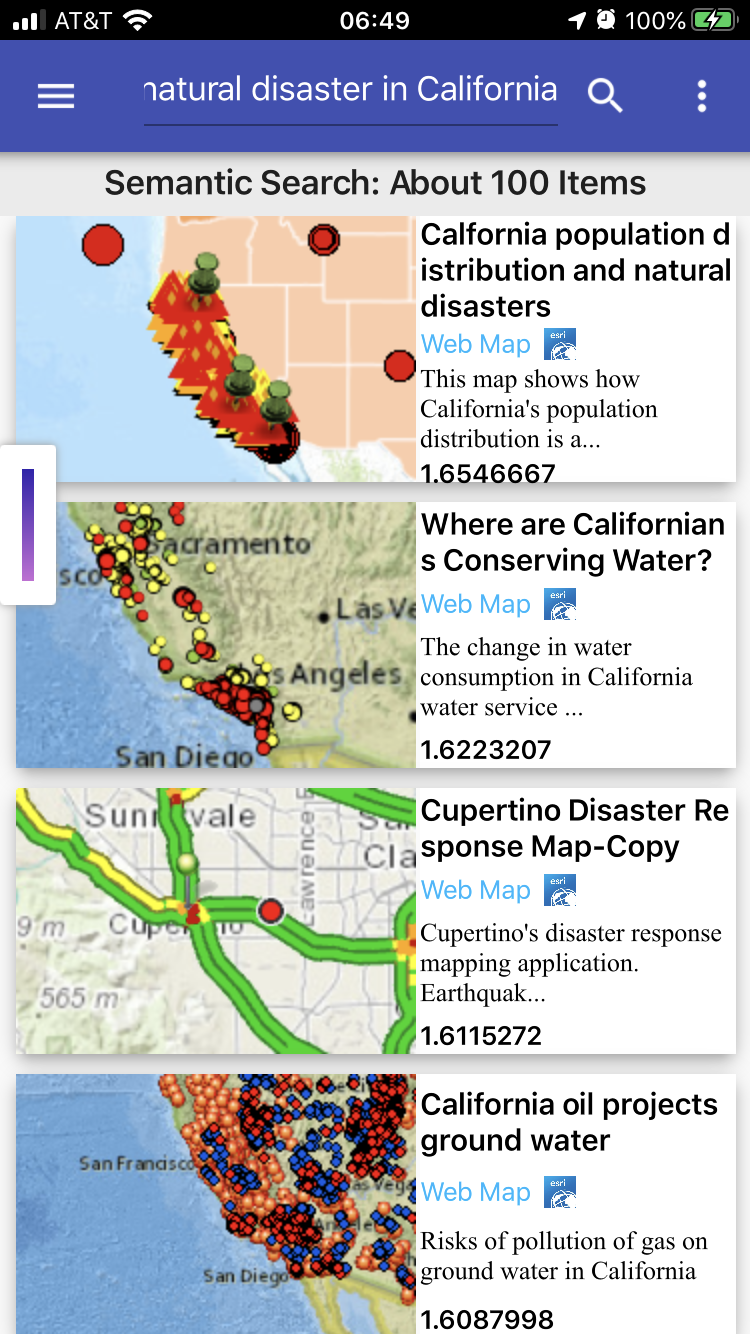}
		\label{fig:mobile_app}
	\end{minipage}
	\caption{
		A mobile application for the semantically-enriched search engine prototype for ArcGIS Online.
	}
	\label{fig:app}
\end{figure}

Based on the presented semantic query expansion framework in Section \ref{sec:method}, we develop a semantically-enriched search engine prototype for ArcGIS Online on top of the established Elasticsearch index. Figure \ref{fig:interface} is a screenshot of the developed system in which the radio buttons \textit{Semantic Search} and \textit{Lucene} correspond to our semantic query expansion based GIR model and the baseline - Lucene's practical scoring function based IR model which we will call it Lucene baseline in the following. This web interface is available through here 
\footnote{\url{http://stko-testing.geog.ucsb.edu:3010/}}
A mobile application is also developed based on AppStudio for ArcGIS (See Figure \ref{fig:app}) .

\subsection{Evaluation}   \label{subsec:eval}
\begin{center}
	\begin{figure}
		\centering
		\includegraphics[width=1.0\textwidth]{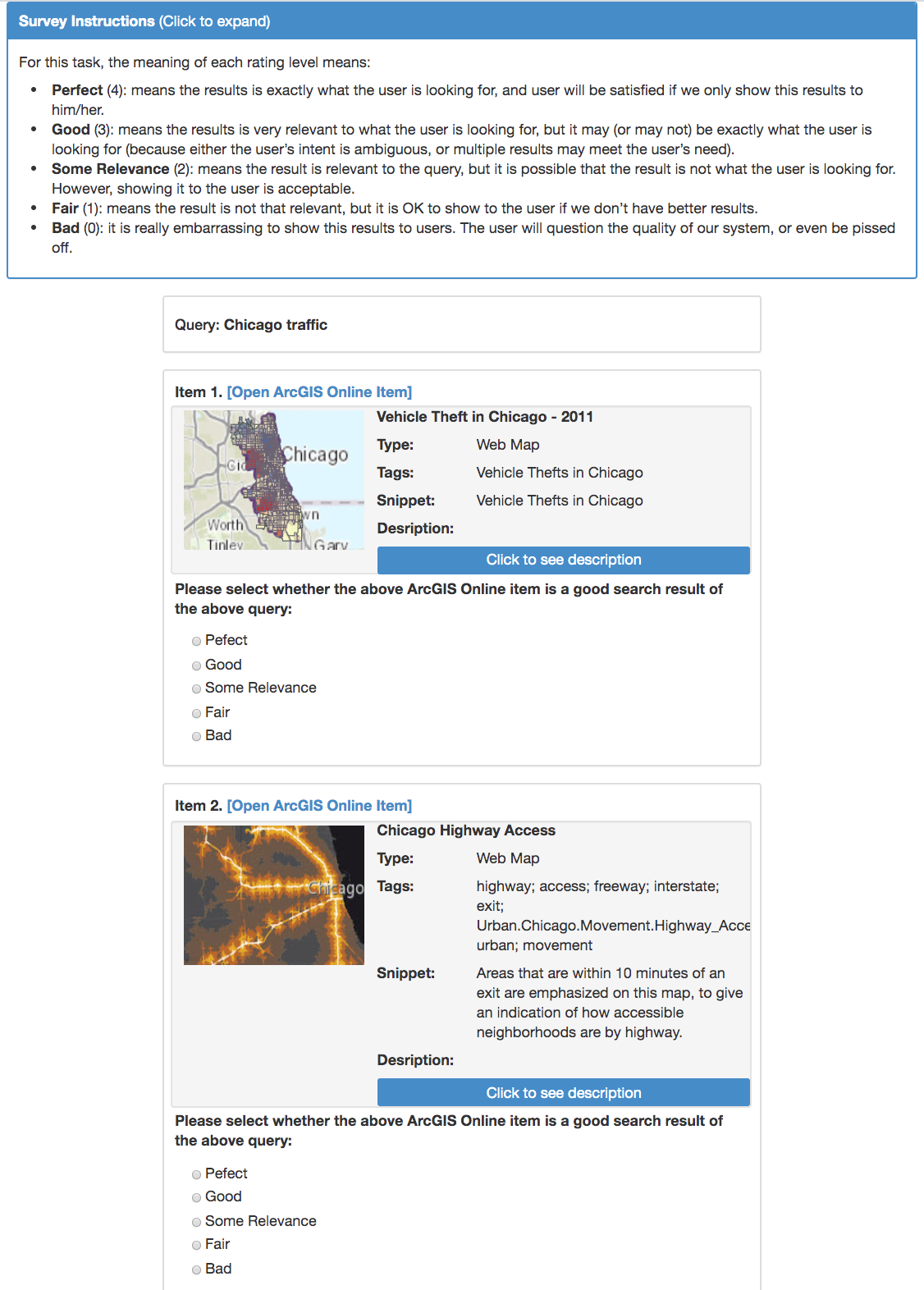}
		\caption{An example of the crowd-facing Web survey form we developed to collect query-item relevance scores.} \label{fig:amt}
	
	\end{figure}
\end{center}

A collection of user search logs is an ideal benchmark dataset to evaluate the presented framework as well as the Lucene baseline as \citet{jiang2018towards} did. As the search logs are not available for the current project, we decide to build our own evaluation dataset. The benchmark dataset construction process can be summarized as follows:
\begin{enumerate}
	\item We collect a query set which consists of 20 queries. All queries can be seen in Table \ref{tab:eval}. The first 10 queries are obtained from \citet{hu2015metadata}, while we manually generate another 10 queries based on the topics and geographic coverage of the collected ArcGIS Online items.
	\item For each query, we get the top 10 search results from our semantic query expansion model as well as the Lucene baseline. 
	\item We create a survey form for each query and each model. Each survey form consists of one query and 10 random ordered ArcGIS Online items. Users are then asked to judge the relevance between the query and each item on an ordinal scale, with labels such as``Perfect'' (4), ``Good'' (3), ``Some Relevance'' (2),``Fair'' (1), and ``Bad'' (0). The numbers in () are used as the corresponding relevance score. An example survey form can be seen in Figure \ref{fig:amt}.
	\item To host these surveys, a crowd-facing Web interface is developed and deployed on Amazon Mechanical Turk sandbox environment.
	\item Eight users completed these surveys who are from different departments of a US university.
\end{enumerate}

In total, we have 40 survey forms, 20 for each GIR model, completed by 8 different accessors. The average relevance score among these 8 accessors' results is treated as the relevance score $rel$ between a query and an item in one form.

\vspace*{-0.7cm}

Discounted Cumulative Gain at top K rank (DCG@K) \citep{carterette2008evaluating,jarvelin2002cumulated} is a typical evaluation metric for information retrieval system. DCG is the weighted sum of ``gains'' of presenting a specific item. The weight is a \textit{discounted} factor by ranking an item at a particular position. For IR systems, DCG at top K rank is defined as shown in Equation \ref{equ:dcg} in which $rel_{i}$ indicates the relevance score between a query and an item, the said gain, and $\dfrac{1}{log_{2} i}$ is the discounted factor based on the current rank $i$.

\begin{equation}
DCG_{K} = rel_{1} + \sum_{i=2}^{K} \dfrac{rel_{i}}{log_{2} i}
\label{equ:dcg}
\end{equation}

We choose DCG@3, DCG@5, and DCG@10 as the evaluation metrics and Table \ref{tab:eval} shows the evaluation results of both our semantic search model and Lucene baseline on each query. Some interesting observations can be made based on Table \ref{tab:eval}:

\begin{enumerate}
	\item By comparing the average DCG scores, our semantic search model outperforms Lucene baseline by a significant margin.
	\item In 17 out of 20 queries, the semantic search model outperforms the Lucene baseline with $\Delta DCG@K > 3$.
	\item As for the two queries (Query 2 and Query 8), the semantic search model provides relatively similar DCG scores ($< 1$).
	\item The only query in which our semantic search model performs clearly worse is Query 10 - \textit{Crimes in Tennessee}. After examining the top 10 search results the two models, we find that:
	\begin{enumerate}
		\item All top 10 search results of Lucene baseline are crime maps about other places such as New York, Miami, or world wide crime reports. Basically Lucene baseline fetches these items based on the thematic similarity.
		\item 9 out of 10 search results of semantic search model are about other topics in Tennessee such as public health, energy, banking while one item is about crimes in neighboring states. As for these 9 items, 7 of them do not contain any place names in their title, snippet, or description but with spatial footprints close to the center of Tennessee. This implies that semantic search model finds these items mostly based on spatial similarity.
		\item There is actually no correct answer about the crime in Tennessee.
		\item However, based solely on these observations we cannot conclude that people pay more attention to thematic similarity than spatial similarity. That is because this bias may be caused by the design of the survey form in which thematic similarity is relatively easy to judge, while spatial similarity is rather difficult as users need to click the link and go to the web map to see the geographic scopes of an item. 
		\item These observations raise an interesting question. How to design an appropriate survey form for evaluating GIR systems in contrast more general IR systems.
	\end{enumerate}
\end{enumerate}

\begin{table}[]
	\caption{Evaluation results of the semantic query expansion model and baseline}
	\label{tab:eval}
	\centering
    \begin{tabular}{c|l|c|c|c|c|c|c|}
\hline
   & \multicolumn{1}{c|}{}         & \multicolumn{3}{c|}{Lucene Baseline}              & \multicolumn{3}{c|}{Semantic Search}              \\ \hline
   & Query                         & \textbf{DCG@3} & \textbf{DCG@5} & \textbf{DCG@10} & \textbf{DCG@3} & \textbf{DCG@5} & \textbf{DCG@10} \\ \hline
1  & New York water                & 1.35           & 1.91           & 4.07            & \textbf{5.90}  & \textbf{8.20}  & \textbf{11.78}  \\ \hline
2  & California fire               & \textbf{9.12}  & \textbf{11.27} & \textbf{14.81}  & 8.25           & 10.67          & 14.36           \\ \hline
3  & California population density & 3.72           & 5.18           & 7.26            & \textbf{6.97}  & \textbf{9.38}  & \textbf{12.88}  \\ \hline
4  & Vacation in Hawaii            & 3.85           & 5.05           & 7.93            & \textbf{8.60}   & \textbf{11.54} & \textbf{15.50}  \\ \hline
5  & Florida flood                 & 8.53           & 10.71          & 13.12           & \textbf{8.70}  & \textbf{10.38} & \textbf{14.60}  \\ \hline
6  & Weather in Iowa               & 3.30           & 5.44           & 7.40            & \textbf{5.51}  & \textbf{7.97}  & \textbf{11.67}  \\ \hline
7  & Chicago traffic               & 6.55           & 7.41           & 10.36           & \textbf{8.81}  & \textbf{11.60} & \textbf{15.55}  \\ \hline
8  & Libraries in Montana          & \textbf{9.40}  & \textbf{12.57} & \textbf{15.30}  & 9.29           & 12.56          & 15.26           \\ \hline
9  & Natural disasters in Utah     & 3.18           & 5.45           & 8.30            & \textbf{7.22}  & \textbf{8.82}  & \textbf{10.85}  \\ \hline
10 & Crimes in Tennessee state     & \textbf{5.03}  & \textbf{7.54}  & \textbf{11.90}  & 1.74           & 1.97           & 2.92            \\ \hline
11 & California transportation     & 5.28           & 5.95           & 7.36            & \textbf{6.44}  & \textbf{8.73}  & \textbf{12.68}  \\ \hline
12 & Agriculture in Michigan       & 6.32           & 7.03           & 8.61            & \textbf{8.69}  & \textbf{9.98}  & \textbf{12.36}  \\ \hline
13 & California weather            & 6.11           & 8.27           & 10.49           & \textbf{6.93}  & \textbf{9.18}  & \textbf{12.42}  \\ \hline
14 & Tourist attraction in LA      & 1.57           & 2.03           & 3.43            & \textbf{6.43}  & \textbf{8.18}  & \textbf{11.35}  \\ \hline
15 & Hurricane in Louisiana        & 4.18           & 5.64           & 9.33            & \textbf{7.11}  & \textbf{9.22}  & \textbf{13.20}  \\ \hline
16 & Universities in Boston        & 2.14           & 2.68           & 4.30            & \textbf{5.66}  & \textbf{7.23}  & \textbf{9.10}   \\ \hline
17 & Hospitals in New York         & 1.90           & 2.63           & 4.41            & \textbf{5.82}  & \textbf{8.70}  & \textbf{12.17}  \\ \hline
18 & Grocery store in Seattle      & 6.12           & 8.17           & 11.28           & \textbf{10.40} & \textbf{13.93} & \textbf{16.99}  \\ \hline
19 & Highways in Los Angeles       & 2.22           & 2.92           & 4.19            & \textbf{7.64}  & \textbf{9.09}  & \textbf{10.26}  \\ \hline
20 & Air pollution of New York     & 6.88           & 8.37           & 9.76            & \textbf{7.04}  & \textbf{9.55}  & \textbf{12.71}  \\ \hline
   & \textbf{Average DCG}          & 4.84           & 6.31           & 8.68            & \textbf{7.16}  & \textbf{9.34}  & \textbf{12.43}  \\ \hline
\end{tabular}
\end{table}

\vspace*{-0.5cm}
\section{Conclusion} \label{sec:conclusion}

In this work, we present a semantic query expansion framework for geographic information retrieval systems. It enriches a user's query from both geospatial and thematic perspectives. Two components are developed for each perspective. By using ArcGIS Online as an example, we develop a semantically enriched search engine prototype by following the proposed query expansion framework. We constructed a benchmark dataset to evaluate the proposed GIR model as well as a widely used baseline model - Lucene's practical scoring function model. The results demonstrate that our semantic query expansion model significantly outperforms the Lucene baseline, thereby highlighting the effectiveness of our proposed approach.

As for future research, we want to improve the efficiency of the presented semantic query expansion framework. We also want to investigate other ways to measure spatial similarity such as Space2Vec \citep{mai2020multiscale}. In addition, we are interested in evaluating the impact of different spatial similarity measures on the performance of GIR systems more generally. Moreover, we plan to investigate the question of whether the added geospatial aspect of GIR will affect the way how we evaluate the system.


\subsection*{Acknowledgments}
This presented work was partially done while the first author was interning at Esri Inc.. This work is partially funded by  Esri Inc. and the NSF award 1936677 C-Accel Pilot - Track A1 (Open Knowledge Network): \textit{Spatially-Explicit Models, Methods, And Services For Open Knowledge Networks}. We thank four Ph.D. students from UC Santa Barbara for evaluation data annotations: Jingyi Xiao, Ning Zhang, Haoxin Zhou, and Yao Xuan. 

\bibliographystyle{agsm}
\bibliography{reference}
\end{document}